\documentclass[twocolumn,nofootinbib,preprintnumbers,superscriptaddress,amsmath,amssymb,showpacs]{revtex4}
\usepackage{graphicx}
\usepackage{dcolumn}
\usepackage{float}
\usepackage{mathptmx, courier, pifont}
\usepackage[scaled=0.92]{helvet}
\usepackage[T1]{fontenc}
\usepackage{textcomp}
\usepackage{color}
\usepackage[normalem]{ulem}

\usepackage[dvipsnames]{xcolor}
\usepackage[colorlinks=true,urlcolor=Blue,linkcolor=Blue]{hyperref}
\usepackage[all]{hypcap}

\begin{document}


\title{Electrostatic-lenses position-sensitive TOF MCP detector for beam diagnostics and new scheme\mbox{} for mass measurements at HIAF}
\thanks{This work was supported by the National Natural Science Foundation of China (Grant No.~11605248, No.~11605249,  No.~11605267, and No.~11805032)}

\author{Jun-hao Liu}
\affiliation{University of Chinese Academy of Sciences,
	Beijing, 100049, People's Republic of China}
\affiliation{Institute of Modern Physics, Chinese Academy of Sciences, Lanzhou, 730000, People's Republic of China}

\author{Zhuang Ge}
\thanks{Corresponding author. Email address: gezhuang@impcas.ac.cn}
\affiliation{University of Chinese Academy of Sciences,
	Beijing, 100049, People's Republic of China}
\affiliation{Institute of Modern Physics, Chinese Academy of Sciences, Lanzhou, 730000, People's Republic of China}

\author{Qian Wang}
\affiliation{University of Chinese Academy of Sciences,
	Beijing, 100049, People's Republic of China}
\affiliation{Institute of Modern Physics, Chinese Academy of Sciences, Lanzhou, 730000, People's Republic of China}

\author{Geng Wang}
\affiliation{University of Chinese Academy of Sciences,
	Beijing, 100049, People's Republic of China}
\affiliation{Institute of Modern Physics, Chinese Academy of Sciences, Lanzhou, 730000, People's Republic of China}

\author{Li-na Sheng}
\affiliation{University of Chinese Academy of Sciences,
	Beijing, 100049, People's Republic of China}
\affiliation{Institute of Modern Physics, Chinese Academy of Sciences, Lanzhou, 730000, People's Republic of China}

\author{Wen-wen Ge}
\affiliation{University of Chinese Academy of Sciences,
	Beijing, 100049, People's Republic of China}
\affiliation{Institute of Modern Physics, Chinese Academy of Sciences, Lanzhou, 730000, People's Republic of China}

\author{Xing Xu}
\affiliation{University of Chinese Academy of Sciences,
	Beijing, 100049, People's Republic of China}
\affiliation{Institute of Modern Physics, Chinese Academy of Sciences, Lanzhou, 730000, People's Republic of China}

\author{Peng Shuai}
\affiliation{University of Chinese Academy of Sciences,
	Beijing, 100049, People's Republic of China}
\affiliation{Institute of Modern Physics, Chinese Academy of Sciences, Lanzhou, 730000, People's Republic of China}

\author{Qi Zeng}
\affiliation{School of Nuclear Science and Engineering, East China University of Technology,  NanChang,  330013,  People's Republic of China}

\author{Bo Wu}
\affiliation{University of Chinese Academy of Sciences,
	Beijing, 100049, People's Republic of China}
\affiliation{Institute of Modern Physics, Chinese Academy of Sciences, Lanzhou, 730000, People's Republic of China}


\date{\today}

\begin{abstract}
A foil–microchannel plate (MCP) detector, which uses electrostatic lenses and possesses both good position and  timing resolutions, has been designed and simulated for beam diagnostics and mass measurements at the next-generation heavy-ion-beam facility HIAF in China. Characterized by low energy loss and good performances of timing and position measurements, it would be located at focal planes in fragment separator HFRS for position monitoring, beam turning, B${\rho}$ measurement, and trajectory reconstruction. Moreover, it will benefit the building-up of a magnetic-rigidity–energy-loss–time-of-flight (B${\rho}$–${\Delta}${E}–TOF) method at HFRS for high-precision in-flight particle identification (PID) of radioactive isotope (RI) beams on an event-by-event basis. Most importantly, the detector can be utilized for in-ring TOF and position measurements, beam-line TOF measurements at two achromatic foci, and position measurements at a dispersive focus of HFRS, thus making it possible to use two complementary mass measurement methods (isochronous mass spectrometry (IMS) at the storage ring SRing and magnetic-rigidity–time-of-flight (B${\rho}$-TOF) at the beam-line HFRS) in one single experimental run. 

\end{abstract}


\maketitle

\section{Introduction}\label{sec.I}
In the last two decades, numerous experiments on nuclear structure and astrophysics studies have been performed using heavy-ion storage rings coupled with in-flight fragment separators~\cite{Lunney, Blaum,sunmz}. These operating ion storage ring facilities are currently located at GSI/Germany~\cite{GSI}, IMP/China~\cite{IMP}, and RIKEN/Japan~\cite{RIKEN}.

In China, the Heavy Ion Research Facility at Lanzhou (HIRFL)~\cite{xia} in the Institute of Modern Physics (IMP) is one major national research facility which focuses on nuclear physics, heavy radioactive isotope (RI) applications, atomic 
physics, and some other interdisciplinary researches. Especially, with the accurate mass values of some very exotic nuclei having been determined using the storage ring CSRe operating as an isochronous mass spectrometer (IMS)~\cite{zhang}, many important results on nuclear astrophysics, such as modeling the rp- and ${\nu}$p-processes to explain the origin of elements and evolution of stars in our universe, have been published~\cite{tuxl,yanxl,xingym}. Because there are still severe limitations when it comes to the intensities of primary beams and the transmission efficiency of CSRe for secondary RIs, a next-generation facility is strongly required.

\begin{figure}[!htb]
\includegraphics
  [width=0.99\hsize]
  {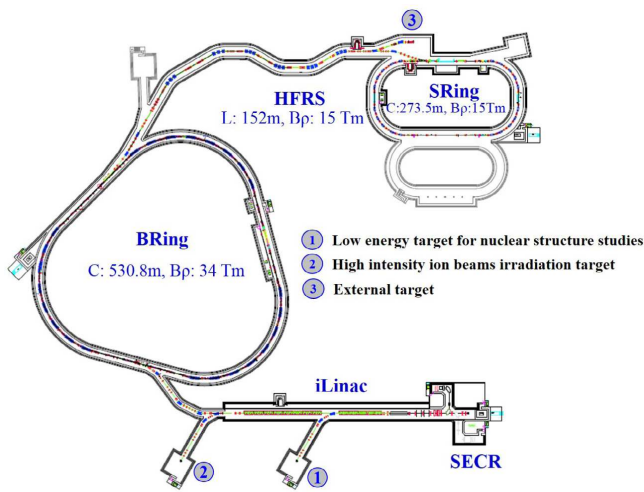}
\caption{Overall schematic layout of the HIAF facilities~\cite{wubo}. }
\label{fig:1}
\end{figure}

Based on the developments and experiences with heavy RI beam accelerators at IMP, a new project, the High Intensity Heavy Ion Accelerator Facility (HIAF)~\cite{wubo}, was proposed in 2009 and was approved by the Chinese government in 2015. The facility is being designed to provide high-intensity, high-quality, and high-energy primary and radioactive ion beams, for the purpose of extending our knowledge on nuclear properties beyond the present limit of existence towards the drip lines and approaching the explosive nuclear-burning processes that occur in astrophysical events. It is envisioned that the secondary beam intensities of the HIAF will be enhanced by about 3–4 orders of magnitude compared to those presently available at HIRFL. The schematic layout of the HIAF project is shown in Fig.~\ref{fig:1}. HIAF consists of a superconducting electron-cyclotron-resonance ion source (SECR), heavy ion Linac (iLinac), booster ring (BRing), HIAF fragment separator (HFRS) (momentum acceptance $\delta$p/p = $\pm$ 2.0 $\times$ 10$^{-2}$), spectrometer ring (SRing), and several experimental terminals.

\begin{figure*}[!htb]
\begin{center}
\includegraphics
  [width=0.9\hsize]
  {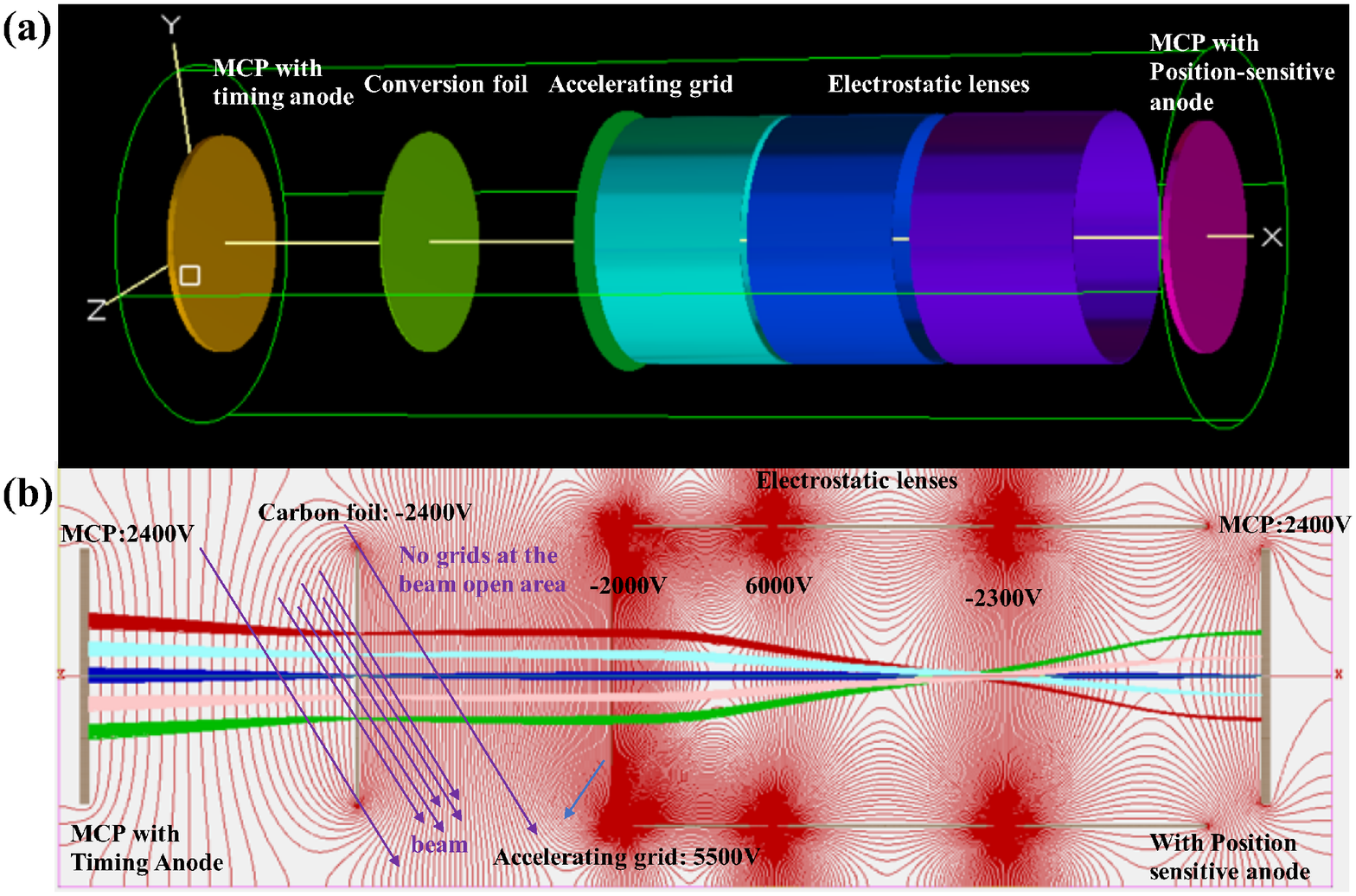}
\caption{(Color online) (a) The structure of the electrostatic-lenses imaging MCP detector.
 (b) The electrostatic potential distribution from SIMION simulation.} 
\label{fig:3}
\end{center}
\end{figure*}
%
Position-sensitive detectors play a crucial role in not only the tuning of the fragment separator HFRS but also the production and delivery of RI beams.
These detectors could be utilized to measure positions and angles (or trajectories) of RI beams at the foci. Through this, beam diagnostics and ion-optical tuning would be made possible. 
In addition, position-sensitive detectors would be used in magnetic rigidity (B${\rho}$) determination for particle identification (PID) of RI beams, wherein trajectory reconstruction is employed to achieve high B${\rho}$ resolution and, hence, excellent PID power. 
Timing detectors are also desired for achieving high-resolution PID of RI beams based on the application of the magnetic-rigidity–energy-loss–time-of-flight ({B${\rho}$}–${\Delta}${E}–TOF) method at the HFRS. Through isochronous mass spectrometry (IMS), such as with ESR/GSI and CSRe/IMP, mass measurements could be performed using in-ring timing detectors~\cite{meib,zhangw} with low energy loss for heavy RI beams at the new storage ring SRing. TOF measurements at two achromatic foci of the high-resolution beam-line HFRS with extra B${\rho}$ measurements for TOF corrections could help realize mass measurements via the well-developed B${\rho}$–TOF method. It is also possible to combine B${\rho}$–TOF mass measurements with IMS mass measurements in the same experimental run, in which the in-ring revolution time is measured simultaneously with the beam-line TOF and B${\rho}$ of each RI. 

For the purposes discussed previously, we have designed a two-segmented foil–microchannel plate (MCP) detector with the following characteristics: (a) very good timing resolution ($<$ 50 ps), (b) two-dimensional position-sensitivity and sub-millimeter resolution, (c) low energy loss and small angular scattering of the heavy ions with the usage of a thin foil, (d) no wires at the passage of the RIs inside the detector, (e) large active area to cover a large beam size, and (f) large detection efficiency.

In the following sections, the structure, principle of timing and position measurements of the newly designed MCP detector, and applications of this type of detector at the next-generation facility HIAF, both on the beam-line HFRS and inside the storage ring SRing, are discussed and presented in detail.
\section{Design of the MCP detector}
\label{sec:structure}
Microchannel plate (MCP) has been widely used as an amplifier in imaging and/or timing detectors at radioactive beam facilities because of its high gain, sub-nanosecond temporal response, low power consumption, stable operation in magnetic fields, sensitivity to a single electron, and compact size~\cite{MCP}.
Foil–MCP detectors, by transporting induced secondary electrons (SEs) from a thin foil toward the MCP surface with different arrangements of the electromagnetic field in order to deduce the timing and/or position information of the heavy ions, are widely utilized in nuclear-physics experiments. Foil–MCP timing detectors have been successfully used in mass-measurement experiments on exotic nuclei in three heavy-ion storage ring facilities: the CSRe/IMP~\cite{meib,zhangw}, ESR/GSI~\cite{GSI-MCP}, and Rare-RI Ring/RIKEN~\cite{nagae,suzuki}, where precise mass measurements via revolution-time measurement with electrostatic-field and magnetic-field crossly-arranged TOF detectors have been successfully performed using the IMS method. Meanwhile, a position-sensitive foil–MCP detector with parallel electrostatic field and magnetic field used for B${\rho}$–TOF mass measurements at NSCL/MSU~\cite{brho-tof} has also demonstrated its high performance and special characteristics for position measurements to deduce the momentum of exotic nuclei at a dispersive focal plane. All the aforementioned described detectors were optimized mostly for timing measurements at the cost of only timing information (ESR and CSRe) or good position resolution, but with relatively worse timing resolution (NSCL) and relatively small detection active areas. Electrostatic mirror-type foil–MCP detectors with good timing performance and medium spatial resolution for the purpose of mass identification studies and beam monitoring have also been developed in many laboratories around the world~\cite{zhuang,e-mcp1,e-MCP}. However, for the electrostatic mirror-type foil–MCP detector, there is a drawback with regard to mirror grids usually being along the passage of the RI beams, which will strongly disturb the circulation of ions when the detector is installed inside a storage ring. To avoid the disadvantages of the detectors described previously, we propose to develop a new concept, a foil–MCP detector with two segmented parts dedicated to performing timing measurement and position determination separately.

The 3D structure of the newly designed detector is shown in Fig.~\ref{fig:3}(a). It consists of a conversion foil, accelerating grid, triplet electrostatic lenses, and two MCPs with a timing anode made of metal~\cite{zhangw,photonis} at the backward side and two-dimensional position-sensitive delay-line anode~\cite{roentdek} at the forward side. The timing anode to be used for the designed detector is from~\cite{photonis}, and the delay-line anode employed is from~\cite{roentdek}. The detector configuration employs a straight structure in which the conversion foil is tilted with an angle of ${\alpha}$ = 30${^{\circ}}$ relative to the heavy-ion beam axis.
Self-sustained carbon foil 5–20 ${\mu}$g/cm${^{2}}$ in thickness or mylar foil coated with aluminum with a thickness of 0.5–2 ${\mu}$m could be used as the conversion foil for SE emission. The accelerating grid equipped with crossed wires with a pitch of 1 mm for both two directions is made of gold-plated tungsten (W+Au) with a diameter of 40 ${\mu}$m.

As shown in Fig.~\ref{fig:3}, the detector has two functional areas: one area is in the backward direction for timing determination, in coincidence with the other area in the forward direction for position measurement.
The backward SEs induced from the foil on impact of a heavy ion are directly accelerated to the MCP via an accelerating potential (\textminus4800 V) between the MCP (biased 2400 V) and the foil (biased \textminus2400 V), and the timing information of the hitting time stamp of the ion is subsequently recorded. Meanwhile, in order to reproduce the position information of the ion on its impact, the emitted forward SEs by the same ion are accelerated by the accelerating grid (biased \textminus2000 V) and focused onto the forward MCP (biased 2400 V) coupled with a position-sensitive anode. The electrostatic lenses could constrain the position dispersion of the SEs and focus the SEs onto the MCP (biased 2400 V) surface, thereby maintaining their position information. The value of the high-voltage supply for each electrode of the triplet-lenses system is indicated in Fig.~\ref{fig:3}b.

One of the most important features of the MCP detector is that the amount of substance along the passage of the RI beams is quite small compared to those of other kinds of position-sensitive gaseous detectors, such as parallel-plate avalanche counters (PPACs)~\cite{PPAC}, multi-wire proportional counters~\cite{MWDC}, and  multi-wire drift chambers~\cite{MWDC}. The amount could be made of approximately 20 ${\mu}$g/cm${^{2}}$  or even less, which is less than 1/100 of the amount for other detectors, thereby perturbing the production and delivery of RI beams significantly less, including when measuring trajectories on an event-by-event basis. Furthermore, the perturbation of the beam is even less because there are no wires   along the passage of RI beams in the MCP detector. Another important feature is that the MCP detectors are durable, with a rate capability up to 1 MHz, and because of their simple structure, maintenance is relatively easy. Components for the construction of the full detector have been acquired and are currently being subjected to testing, with results expecting to be published in the near future~\cite{zhuang-anode}.
\begin{figure}[!htb]
%
\includegraphics
  [width=0.95\hsize]
{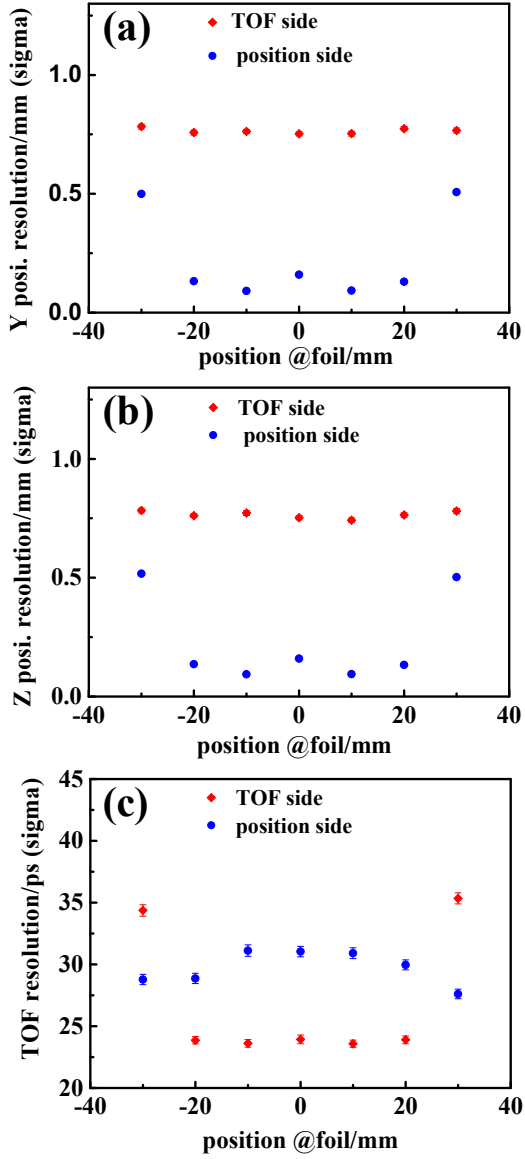}
\caption{(Color online) Comparison of (a) Y-direction, (b) Z-direction position resolution, and (c) timing resolution as a function of the position at the foil for the timing side (depicted as TOF side in legend) and position-sensitive side (depicted as position side in the legend), respectively. These results are from simulation via SIMION~\cite{SIMION}.} 
\label{fig:4}
\end{figure}
\section{Simulation}
\label{sec:simulation}
To optimize the design and performance of the MCP detector, a simulation of the transport of SEs induced from the conversion foil is performed using the SIMION software package~\cite{SIMION}. 
In this simulation procedure, SE energies were assigned to a range of 0–20 eV with a mean value $\sim$2 eV, which corresponds to the main-component electrons emitted from the foil with a percentage of 85 \%~\cite{electron}. The emission angles of the SEs lie between \textminus90 and +90 degrees relative to the foil surface in both forward and backward directions. The main-component SEs~\cite{electron} corresponding to the low-energy electrons, with a percentage of 85\% among all the SEs, will be guided to the MCP surface for detection by the electrostatic field provided by the potential plates. Studying these electrons induced from the thin conversion foil could therefore provide information on the incident ions.
Each parameter of the uniformly distributed electrons is generated using a random generator in SIMION. The distribution of the simulated electrostatic field in the XZ plane view is shown in Fig.~\ref{fig:3}(b). The trajectories of grouped SEs induced by heavy RIs at 5 points on the foil are illustrated with corresponding colors. The simulation at each of the 5 points was performed using 10,000 primary electrons.

As shown in Fig.~\ref{fig:4}, the results are extracted from the SIMION simulation. The obtained position resolution and timing resolution are demonstrated for both sides, which are depicted as TOF side and position side in the legends of Fig.~\ref{fig:4}. The Y- and Z-coordinate position resolutions of the MCP detector at both timing and resolution-sensitive sides as a function of the corresponding positions on the foil are shown in Fig.~\ref{fig:4}(a) and (b). In Fig.~\ref{fig:4}(c), the timing resolutions of the position-sensitive side and dedicated timing side as a function of the corresponding positions on the foil are demonstrated. The intrinsic timing resolution of the timing anode is normally < 20 ps (from our previous test in~\cite{zhangw}), and the  intrinsic timing resolution of the position-sensitive delay-line anode is about 25 ps~\cite{roentdek}, with correction of position dependence for the timing. To avoid influences of the large intrinsic timing resolution from the delay-line anode on the resulting TOF resolution and to achieve as high a timing resolution as possible, we employ the timing anode dedicatedly for the TOF measurement side. The intrinsic position resolution of $\sim$0.1 mm~\cite{roentdek} for the position-sensitive delay-line anode can nearly be ignored compared to the position resolution of 0.5–1 mm for the position-sensitive side of the detector.
The desired position resolution and timing resolution for this type of detector are $\leq$ 1 mm (in ${\sigma}$ for two dimensions) and 20–40 ps (in  ${\sigma}$), respectively, with an effective area 60 mm in diameter. With the timing measurement of each ion's impact on the foil in addition to its position, the correction of position dependence for the timing could be realized in order to achieve high performance for the timing measurement, independent of its position. The effective area of this type of detector would be increased to be as large as 100 mm in diameter in the future design.
\section{Application of the detector at HIAF}
\subsection{Utilization for the HFRS}
\label{sec:HFRS}
\begin{figure*}[!htb]%
\includegraphics
   [width=0.85\hsize]
   {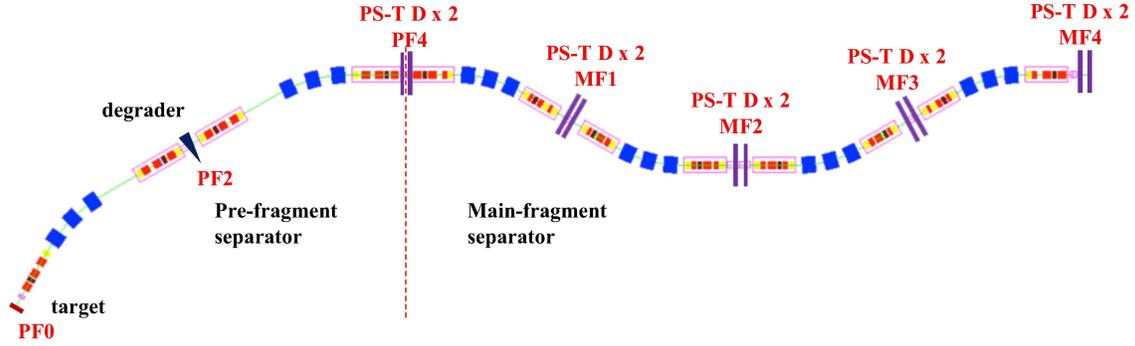}
\caption{(Color online) The completion status of installation of the foil–MCP detectors at HFRS. Dual foil–MCP detectors indicated in purple blocks could be employed for position, angle, and timing of arrival measurements of RIs at foci (PF4, MF1–4) of HFRS event by event. The ``PS-T D'' denotes position-sensitive timing detector.}
\label{fig:5}
\end{figure*}

\begin{figure*}[!bht]%
\includegraphics
   [width=0.85\hsize]
   {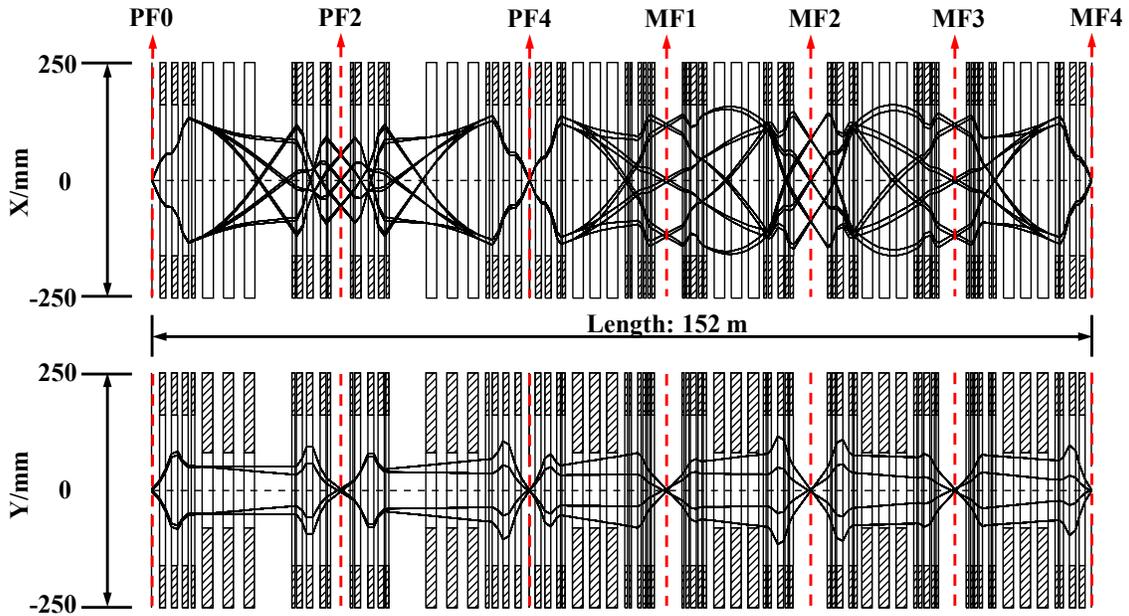}
\caption{(Color online) Ion-optical design of the HFRS. High-resolution ion-optics (first-order) mode of the HFRS system with achromatic foci at PF0, PF4, and MF4, dispersive foci at PF2, MF1, MF2, and MF3 in the upper (horizontal, denoted with X) plane, and  achromatic foci at PF0, PF2, PF4, and MF1–MF4 in the lower (vertical, denoted with Y) plane, respectively.} 
\label{fig:2}
\end{figure*}

HFRS is a long-fragment separator consisting of two functional parts: the pre-fragment (PF) separator for selection and separation of RIs and the main-fragment (MF) separator for high-resolution PID of RIs. The HFRS is primarily a powerful separator but can also be operated as a high-resolution spectrometer simultaneously. A schematic view of HFRS and proposed completion status of foil–MCP detectors (demonstrated in purple at each focus) at the HFRS are shown in Fig.~\ref{fig:5}. A high-resolution ion-optical design of HFRS is shown in Fig.~\ref{fig:2}.

An achromatic system normally has the best performance for spatial separation of mono-isotopic beams. The ion-optical system of the HFRS is doubly achromatic at the final focal plane. Thin production targets could be employed and installed at PF0. A degrader at PF2 is shaped such that the achromatism is preserved at the final focal plane MF4. HFRS will be employed as a two-stage separator: the first stage, from PF0 to PF4, is used for separation of the nuclei of interest through a B${\rho}$–$\Delta$E–B${\rho}$ selection, and the second stage, from PF4 to MF4, is for PID of the RIs via a B${\rho}$–$\Delta$E–TOF method. The second stage is also used to deduce velocity and momentum dispersions of RIs on an event-by-event basis.

The trajectory of an ion in a magnetic field {B} depends on its mass number A, charge Q = Ze (Z is the proton number, and e is the charge of electron), and momentum P and can be described by the equation:
\begin{equation}
B\rho= \frac{P}{Q} = \frac{ mv \gamma} {Q}= \frac{A}{Z}\frac{cm_{u}}{e} \gamma \beta,
\label{eq:brho}
\end{equation}
where m$_{u}$ $\approx$ 931.494 MeV is the atomic mass unit, c is the speed of light, ${\rho}$ is the radius of curvature, $\beta$ = v/c is the  relativistic factor, and $\gamma$ = (1 \textminus ${\beta^{2})^{-1/2}}$ is the relativistic Lorentz factor. The secondary nuclei are produced by fragmentation at PF0, and their velocity is nearly constant. The selection in B${\rho}$ with the dipoles between PF0 and PF2 is then equivalent to a selection in A/Z. 
The identification of the resulting secondary cocktail beam on an event-by-event basis is performed in the second part of the HFRS. 
The nuclei of interest are transmitted up to the doubly achromatic MF4 focal plane (a beam focused in both the horizontal and vertical planes with no dependence on angle and momentum). 

The PID of the secondary cocktail beams is achieved using the B${\rho}$–$\Delta$E–TOF method in the HFRS spectrometer, which ensures tagging of every particle delivered in the form of a cocktail beam from projectile or in-flight fission fragments.

The PID at the secondary stage, from PF4 to MF4, provided the atomic number Z and mass-to-charge ratio A/Q of a fragment by measuring the energy loss, magnetic rigidity, and time of flight ($\Delta$E–B${\rho}$–TOF) with the corresponding beam-line detectors. The B${\rho}$ and TOF of the RIs could be measured with relatively high enough position and timing resolutions using the MCP detectors designed in this study. The energy loss of RIs can be measured by an ionization chamber (IC).
The TOF between two foci (PF4 and MF4) could be measured by the corresponding MCP detectors, where TOF = L/($\beta$ c), which can be used for velocity (${\beta}$) measurements with a correction via precise B${\rho}$ measurement by an MCP detector at the focus MF2.
From  B$\rho$= $\frac{P}{Q}$ = $\frac{ mv \gamma} {Q}$= $\frac{A}{Q}\frac{c m_{u}}{e} \gamma \beta$, we can get :
\begin{equation}
\frac{A}{Q} = \frac{B\rho}{\beta \gamma} \frac{e}{m_{u}c}.
\label{aoq-cal}
\end{equation}
The atomic number Z of a heavy ion is deduced based on the energy loss of the charged particle in the IC. The energy loss of a heavy ion in a unit thickness of material can be described using the Bethe–Bloch formula~\cite{wiki,bethe1,bethe2}, which is defined as:
\begin{equation}
 -\frac{dE}{dX}
=\frac{4\pi Z^2 n_{m}}{m_{e} c^2 \beta^2}{(\frac{e^{2}}{4\pi\varepsilon_{0}})}^{2}[ln{\frac{2m_{e} c^2\beta^2}{I} } - ln(1-{\beta}^{2}) - {\beta}^{2}], 
\label{bethe-bloch-dedx}
\end{equation}
where  n$_{m}$ = ({N$_{A}$Z$_{m}$$\rho_{m}$})/({A$_{m}$M$_{u}$}) is the electron density of the material (counter gas).
The formula for calculating the atomic number Z can then be written as:
\begin{equation}
Z = a_{1}\beta\sqrt{\frac{\Delta E}{ln(\frac{2m_{e}c^{2}\beta^{2}}{I}) - ln(1-\beta^{2})- \beta^{2}}} +a_{2}, 
\label{z-calibration}
\end{equation}
where a${_{1}}$ and a${_{2}}$ are parameters to be calibrated from the experiment in order to reconstruct the Z value for each ion, I is the mean excitation potential of the counter gas, Z$_{m}$ is the atomic number of the counter gas, $\rho_{m}$ is the density of the counter gas, A$_{m}$ is the relative atomic mass of the counter gas, N$_{A}$ is the Avogadro number, M$_{u}$ is the molar mass constant, Z is the atomic number of heavy ion, m$_{e}$ = 0.511 MeV is the electron mass, and $\varepsilon_{0}$ is the vacuum permittivity.

The variable ${\Delta}$E in Eq.~\eqref{z-calibration} is from the  energy losses in the IC, while the velocity ${\beta}$ of the heavy ion in the IC could be derived from the TOF measured with the MCP detectors between the PF4 and MF4 focal planes. The position-sensitive detector in the focal plane MF2 of the HFRS with high momentum resolution can be used to determine the B${\rho}$/${\beta}$ spread of each fragment before it is injected into the storage ring SRing. The precise B${\rho}$ determination in Eq.~\eqref{aoq-cal} is achieved by means of trajectory reconstruction, in which measured particle trajectories are combined with ion-optical transfer matrix elements deduced from experimental data from the MCP detectors. A correlation of measured beam trajectories at the initial and final focal planes provides us with direct information on the ion-optical transfer-map elements. For example, a correlation between the final position (MF4 x) and the initial angle (PF4) provides us with a measurement of the (x$|$a) element of the transport map. 

In addition to high-resolution PID of RIs, a fast-response tuning method could be established with beam-line trajectory monitoring and deduction of ion-optical transfer matrix elements, to help overcome the difficulties caused by low intensity, and large longitudinal and transverse emittances of the exotic RI beams.

\subsection{Utilization for a new scheme of mass measurements at HIAF}
\label{sec:new-scheme}
Strong interest in fast, high-accuracy, and high-precision mass measurements for exotic nuclides, due to their importance in nuclear astrophysics and nuclear-structure studies, has triggered the development of a variety of techniques for mass measurement around the world. These techniques include: the Penning trap mass spectrometers ISOLTRAP/ISOLDE~\cite{Trap}; Schottky mass spectroscopy (SMS) or isochronous mass spectroscopy (IMS) at the storage rings ESR/GSI~\cite{zhang}, CSRe/IMP~\cite{zhang}, and Rare-RI Ring/RIKEN~\cite{R3}; beam-line time-of-flight (TOF) measurements implemented at several facilities, including the SPEG spectrometer at GANIL~\cite{SPEG}, TOFI spectrometer/LANL~\cite{TOFI}, and S800 Spectrograph/NSCL~\cite{brho-tof}; and multi-reflection time-of-flight mass spectrometer (MR-TOF MS) at ISOLDE~\cite{MR-TOF}.

For HIAF, with the development of high-performance position-sensitive and timing detectors, the use of IMS method at the storage ring SRing and B${\rho}$–TOF method at the HFRS could be established. For in-ring mass deduction and beam-line B${\rho}$-TOF mass measurements, low energy loss and energy straggling of RIs in position-sensitive and timing detectors during their detection is indispensable for momentum/velocity reconstruction with high accuracy and precision. With the development of the foil–MCP detector for precise and simultaneous position and timing measurements, as described in section~\ref{sec:structure}, it would be possible to establish a new scheme for mass measurements at the HIAF, in which two complementary methods, B${\rho}$–TOF mass measurements at the beam-line HFRS and IMS mass measurements at the storage ring SRing, could be carried out in the same experimental run.

\subsubsection{Mass measurements by  ${B{\rho}}$–TOF method on HFRS}
\label{sec:brho-tof}
The B${\rho}$–TOF technique for mass measurements of exotic nuclei is based on measurements of the B${\rho}$ and corresponding TOF at the beam-line with a length of L for the ion, such that:
\begin{equation}
B{\rho} = \frac{\gamma{m}}{q}\frac{L}{TOF}.
\label{magnetic-rigidity}
\end{equation}
 The mass-to-charge ratio (m/q) is derived from Eq.~\eqref{magnetic-rigidity} as:
 \begin{equation}
 m/q = \frac{B{\rho}}{\gamma{L/TOF}} = B{\rho}\sqrt{(\frac{TOF}{L})^{2} - (\frac{1}{c})^{2}},
\label{magnetic-rigidity-mass}
\end{equation}
where c is the speed of light, and ${\gamma}$ is the Lorentz factor. TOF could be determined with very high precision using the MCP detectors at PF4 and MF4, as shown in Fig.~\ref{fig:5}, while the measurements of B${\rho}$ at MF2 and flight length L from PF4 to MF4 would severely limit the precision of the resultant mass. Thus, in practice, nuclei with well-known masses are measured together with mass-unknown nuclei in order to calibrate the relationship between TOF, B${\rho}$, and mass. 
The derived mass uncertainty from Eq.~\eqref{magnetic-rigidity-mass} is defined as:
\begin{equation}
\frac{\delta m}{m} = \sqrt{ \frac{\delta{(B{\rho})}^{2}}{(B{\rho})^{2}} + \frac{1}{\left[1 - \frac{L^{2}}{{(TOF)}^{2}{c}^{2}} \right]^{2}}\left(\frac{\delta (TOF)^{2}}{{(TOF)}^{2}} + \frac{{\delta (L)}^{2}}{L^{2}}\right)}.
\label{magnetic-rigidity-mass-error}
\end{equation}
The mass resolving power of this technique strongly depends on the MCP detector (TOF from PF4 to MF4 and position measurement at MF2) resolutions and the momentum resolution of the beam-line from the ion-optical design. The path length between PF4 and MF4 is $\sim$100 m. With a typical TOF of $\sim$600 ns from PF4 to MF4, a TOF resolution better than 60 ps from the MCP detectors at PF4 and MF4 will result in a TOF resolution that is better than 10$^{-4}$. With a typical designed momentum resolving power of $\sim$100 mm/\% for the beam-line, we need position measurement at the dispersive focus MF2 with a resolution better than 1 mm (in $\sigma$, FWHM $\sim$2.35 mm) to achieve a resolution of B${\rho}$ measurement $\sim$10$^{-4}$. Because of the beam tracking at PF4 and MF4 with position resolutions better than 1 mm (in $\sigma$), the uncertainty of path length ($\sim$ 100 m) reconstruction contributing to the mass uncertainty will be better than 10$^{-5}$, which can be ignored compared to other uncertainties, such as uncertainties of B${\rho}$ measurements ($\sim$10$^{-4}$) and TOF measurements ($\sim$10$^{-4}$). The MCP detector designed with a position resolution better than 1 mm and timing resolution of 20–40 ps will help realize the uncertainties of TOF, B${\rho}$, and path length L measurements at better than 10$^{-4}$, 10$^{-4}$, and 10$^{-5}$, respectively. The final mass resolving power for B${\rho}$–TOF method will be $\sim$10$^{4}$ when all the contributing factors in Eq.~\eqref{magnetic-rigidity-mass-error} are considered.  
B${\rho}$–TOF mass measurement with a typical TOF $\sim$1 ${\mu}$s provides mass data with an accuracy from $10^{-4}$ up to the level of $10^{-6}$, which usually depends on the statistics, and allows simultaneous measurements of many nuclides, reference isotopes, and isotopes of interest. For example, if we accumulate counts of 10,000 for one species of ion with a mass resolving power of $10^{4}$, a statistical mass uncertainty of $10^{-6}$ will be achieved; if one event is recorded, then the statistical mass uncertainty will be $10^{-4}$.
\subsubsection{Storage-ring mass spectrometry with SRing}
\label{sec:IMS}
Storage-ring spectrometers are powerful devices for high-accuracy (10$^{-6}$–10$^{-7}$) mass measurements of very short-lived isotopes. With regard to the measurements of momentum/velocity techniques, new developments and efforts, such as the in-ring double-TOF method used at CSRe/IMP and CRing/GSI and the additional beam-line B${\rho}$ measurement coupled with individual injection used at Rare-RI Ring/RIKEN, have extended the applicability and capability of storage-ring mass spectrometry. 

The SRing, which is an essential part of the HIAF, is a dedicated storage ring for precision mass and half-life measurements of short-lived nuclei with lifetimes down to several tens of microseconds.
The basic principle for storage-ring mass spectrometry describing the relationship between mass-over-charge ratio (m/q) and revolution period (T) or revolution frequency (f) can be quantitatively expressed as a first-order approximation:
\begin{equation}
\frac{d T}{T}  
 = - \frac{d f}{f} = \frac{1}{{\gamma_{t}}^{2}} \frac{d (m/q)}{m/q}-\left( 1-\frac{\gamma^2}{\gamma_t^2} \right)\frac{d v}{v},\label{eq0}
\end{equation}    
where ${\gamma_t}$ is the so-called transition energy of the ring.
In order to determine the m/q values with the measured revolution frequencies, or alternatively, with the revolution times of the ions, the second term of the right-hand side of Eq.~\eqref{eq0} has to be made negligibly small: 
\begin{equation}
( 1-\frac{\gamma^2}{\gamma_t^2})\frac{d v}{v}  \rightarrow 0.
\label{eq3}
\end{equation}
Based on this principle, two complementary experimental methods to satisfy Eq.~\eqref{eq3}, namely Schottky (SMS) and isochronous (IMS) mass spectrometry, have been developed for accurate mass measurements~\cite{zhang}.
One way is to cool the ions via stochastic and electron cooling, which forces all stored ions toward the same mean velocity, thereby reducing the velocity spread to roughly 5 $\times$ 10$^{-7}$ for a low-intensity beam. 
Another way is to use a special ion-optical setting of the ring and inject the ions with $\gamma$ = $\gamma_t$, which is the basis for IMS. 
Exotic nuclei with half-lives shorter than the cooling time can be studied by operating the storage ring in the isochronous ion-optical mode.

In the conventional IMS method used at ESR/GSI or CSRe/IMP, one TOF detector is installed in the ring for particle identification via measurement of the in-ring TOF spectrum and deduction of mass without extra B${\rho}$ or velocity measurements of ions. In the case of Rare-RI Ring, the revolution time is deduced from the start TOF detector before injection and stop TOF detector at the extraction line.

To decrease the spread of revolution period, additional B${\rho}$ or ${\beta}$ measurements of stored ions should be precisely measured.
It is assumed that ions with the same magnetic rigidity will move around the same closed orbit, regardless of their species. Therefore, correction of magnetic rigidity can be established via the correction of the corresponding orbit. 
By employing this additional information, the mass resolving power of IMS will be significantly improved ~\cite{shuai,xiangchen}.

\begin{figure}[htb!]  
\includegraphics  
[width=0.85\hsize]
{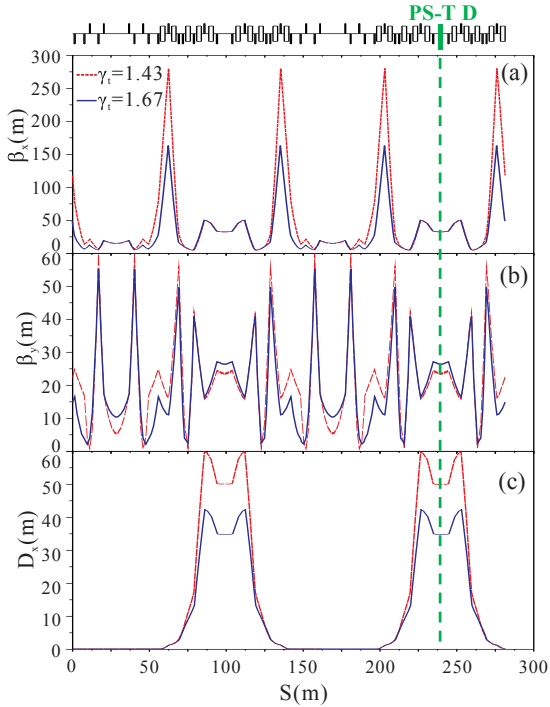}
\caption{(Color online)  Ion-optical functions of the SRing in isochronous mode: (a) horizontal beta function ${\beta_{x}}$, (b) vertical beta function ${\beta_{y}}$, and (c) horizontal dispersion function {D$_{x}$}. S is the path length along the SRing. The red and blue lines indicate the ion-optical modes with ${\gamma_{t}}$ = 1.43 and 1.67, respectively. A foil–MCP detector denoted by ``PS-T D'' in green is installed at a dispersive arc of SRing.}  
\label{fig:7}
\end{figure}

In IMS mass-measurement experiments at GSI, where the high resolution of the FRS is used to determine the B${\rho}$ of the injected fragments within $\sim$10$^{-4}$ at the second dispersive focal plane via a modified slit system, the mass resolution achieved with this B${\rho}$-tagging method is significantly improved together with the time spectra resolution~\cite{Geissel}.
Another method, known as the double-TOF method, is to measure the velocity of a heavy ion circulating in a storage ring event by event, by using two TOF detectors installed at the straight section of the storage ring. 
The precision measurement of velocity with double-TOF detectors has been developed at the straight section of CSRe. The Collector Ring (CR) at FAIR and the SRing at HIAF will both be equipped with double-TOF detectors.

The SRing is proposed to measure the exotic nuclei in IMS mode with additional velocity or magnetic rigidity measurement for TOF correction of non-isochronous RIs. Therefore, the correction of the in-ring revolution time by B${\rho}$ (equivalent to ${\beta}$ for a certain close orbit) measurements with double-TOF detectors can be carried out. Alternatively, to realize the correction of non-isochronous TOF of RIs in SRing, a foil–MCP detector with a simultaneously high position and timing performance, as designed in this study, could be installed at a dispersive arc section of the SRing to measure the ion's position in order to deduce the B${\rho}$ of each ion event by event. As shown in Fig.~\ref{fig:7}, the values of horizontal beta function  ${\beta_{x}}$, vertical beta function ${\beta_{y}}$, and horizontal  dispersion function {D$_{x}$} along the path length are demonstrated for SRing in isochronous mode. Two different ion-optical modes with ${\gamma_{t}}$ = 1.43 and 1.67 are calculated. A foil–MCP detector denoted by ``PS-T D'' in green is installed at a dispersive arc of the SRing where the value of horizontal dispersion function {D$_{x}$} is large, in order to deduce the  B${\rho}$ value precisely. The maximum values of dispersion function in the arc where the MCP detector is located reach 35 m and 50 m for ${\gamma_{t}}$ = 1.43 and 1.67 cases, respectively. 
Compared to the double-TOF method, in principle, only one foil–MCP detector with relatively lower energy loss of the circulating RIs can realize the same correction effect. The double-TOF detectors can actually be replaced by one compact detector with high performance in terms of both good timing and position measurements, in order to reduce accumulated energy losses of RIs when passing through the foil of the detector for hundreds to thousands of turns.

The mass-to-charge ratio m${_{1}}$/q${_{1}}$ for a nucleus with unknown mass could be deduced from known mass  m${_{0}}$/q${_{0}}$ with extra B${\rho}$/${\beta}$ measurements and can be expressed as:
\begin{widetext}
\begin{equation}
 \frac{m_{1}}{q_{1}} 
 =  \frac{m_{0}}{q_{0}} \frac{T_{1}}{T_{0}}\sqrt{\frac{1-\beta_{1}^{2}}{{1-{(\frac{T_{1}}{T_{0}})^{2}\beta_{1}}}^{2}}}
  =   \frac{m_{0}}{q_{0}} \frac{T_{1}}{T_{0}}\sqrt{\frac{1-(\frac{T_{0}}{T_{1}})^{2}}{{{\left( {\frac{m_{0}}{q_{0}}c}{{(B\rho)_{0}}^{-1}}\right) }}^{2}}+1},
\label{orbit-IMS-vel}
\end{equation}
\end{widetext}
where m is the mass of the ion, q is charge of the ion, B${\rho}$ is the magnetic rigidity, and B is the magnetic field of the ring. {T$_{0}$} and {T$_{1}$} are revolution times of an ion with known mass, and the relativistic factor $\beta$ = v/c denotes the particle velocity relative to the velocity of light c. When a particle of interest with a mass-to-charge ratio {m$_1$}/{q$_1$} has the same momentum as that of a reference particle with a mass-to-charge ratio {m$_{0}$}/{q$_{0}$} (assuming mass is known with a precision better than 10$^{-6}$), the flight path length of these particles become identical in the isochronous storage ring. {m$_{1}$}/{q$_{1}$} can be deduced with a precision of 10$^{-6}$ from the ratio m${_0}$/q${_0}$ of the reference nucleus using the revolution time (T), with a precision of 10$^{-6}$, and velocity (${\beta}$)/momentum (B${\rho}$) measurements, with a precision of 10$^{-4}$, of all ions.
The parameters K = $\sqrt{{(1-\beta_{1}^{2})}/{({1-{(\frac{T_{1}}{T_{0}})^{2}\beta_{1}}}^{2})}}$
and 
P = $\sqrt{{(1-(\frac{T_{0}}{T_{1}})^{2})}/{{{( {\frac{m_{0}}{q_{0}}c}{{(B\rho)_{0}}^{-1}}) }}^{2}}+1}$
are referred to as velocity correction factor and momentum correction factor, respectively.
The circumference of SRing is 273.5 m, with the radius ${\rho}$ of the dispersive arc much larger than 10 m~\cite{wubo}. With a position resolution better than 1 mm for the MCP detector, the resolution of B${\rho}$ measurement (equivalent to $\beta$ measurement) will be much better than 10$^{-4}$, which is the required resolution to achieve a mass resolution of 10$^{-6}$ for the in-ring mass measurements. With a revolution time T $\sim$1 $\mu$s in SRing, an ion possessing a circulating turn number (N) larger than 300 will have an uncertainty of the revolution time: $\delta$T = $\sqrt{2}$$\delta T_{MCP}$/(T$\times$N), where $\delta T_{MCP}$ is the timing resolution of the MCP detector.  The resolution of the MCP detector ($\delta T_{MCP}$)  $\le$ 100 ps is enough to realize the uncertainty of the revolution time to be better than 10$^{-6}$. Therefore, the MCP detector designed with an expected timing resolution of 20–40 ps and position resolution better than 1 mm could help realize the scheme of mass measurements discussed previously.

In summary, B${\rho}$–TOF method would be a perfect match with multi-physics experiments in the next-generation RI beam facility HIAF in China.
To reach the most exotic area of the chart of nuclides and to have an effective use for the precious beam time, two complementary TOF techniques for mass measurements can be employed at HIAF in one experimental run. One is the well-developed B${\rho}$–TOF method at the beam-line HFRS, and the other is the IMS method at the SRing with extra B${\rho}$ (or ${\beta}$) measurements event by event. 
\section{Summary and conclusion}
\label{sec:4}
In this paper, we outlined the design and development of a new type of MCP detector at the next-generation facility HIAF, including its operating principles, design and specifications, characteristics, and performance via simulation.
This type of detector is a versatile instrument which can be used on the beam-line HFRS for two-dimensional position measurement to reconstruct beam trajectory for the purposes of beam tuning, high-resolution PID, beam-line momentum measurements of heavy ions for velocity reconstruction, and beam-line TOF measurements between two foci to ensure high-resolution PID and to deduce the velocity of each RI. Meanwhile, it could also be directly used for position monitoring and revolution-time measurement turn by turn inside the storage ring SRing for mass measurements. A new scheme for mass measurement, in which two complementary methods, IMS and B${\rho}$–TOF, would be performed simultaneously in one experimental setting using this type of detectors, has been proposed for HIAF. A single in-ring TOF detector with position-sensitivity to deduce the B${\rho}$ of circulating RIs for IMS mass measurements, which has an advantage equivalent to that of the double-TOF method but with less energy loss for the RIs during their passage, has also been proposed in this paper. The detector will be constructed and tested before the commissioning run of the HIAF.

Through the promising new scheme for mass measurement at the next-generation facility HIAF, we could measure the masses of the most exotic nuclei closely approaching the drip lines that are far away from the valley of stability, which have previously been difficult to study because of their low production rates. In particular, with the foreseen high-accuracy and high-precision mass values from this project as significant inputs in modelling calculations for explosive nucleosynthesis processes (such as r-, rp- and ${\nu}$p-processes), we could help throw light on the origin of elements and evolution of stars in our universe.


\end{document}